\documentclass[twocolumn,aps,showpacs,prl]{revtex4}
\usepackage{epsfig}
\usepackage[english]{babel}
\usepackage{latexsym}
\usepackage{graphics}
\usepackage{subfigure}

\begin{document}

\title{Berry Phase Effects on Dynamics of Quasiparticles in a Superfluid
with a Vortex}
\author{Chuanwei Zhang$^{1,2}$, Artem M. Dudarev$^{1,3}$, and Qian Niu$^{1}$}

\begin{abstract}
We study quasiparticle dynamics in a Bose-Einstein condensate with a vortex
by following the center of mass motion of a Bogoliubov wavepacket, and find
important Berry phase effects due to the background flow. We show that Berry
phase invalidates the usual canonical relation between the mechanical
momentum and position variables, leading to important modifications of
quasiparticle statistics and thermodynamic properties of the condensates.
Applying these results to a vortex in an infinite uniform superfluid, we
find that the total transverse force acting on the vortex is proportional to
the superfluid density. We propose an experimental setup to directly observe
Berry phase effects through measuring local thermal atoms momentum
distribution around a vortex.
\end{abstract}

\affiliation{$^1$ Department of Physics, The University of Texas, Austin, Texas
78712-1081 USA\\
$^{2}$ Condensed Matter Theory Center, Department of Physics, University of
Maryland, College Park, Maryland 20742 USA\\
$^{3}$ Max-Planck-Institut f\"{u}r Physik Komplexer Systeme, Dresden, 01187
Germany}
\maketitle

Quasiparticles in a superfluid play an essential role in its thermodynamic
as well as dynamic properties \cite{Landau}. Owing to the realization of
Bose-Einstein condensates (BEC) with ultra-cold atoms, much detailed
investigation of the quasiparticles becomes possible \cite{review,BGS}. In
the past several years, extensive experimental and theoretical work has been
devoted to the study of quasiparticle dynamics in static BECs \cite%
{BGS,Clark,Csordas,niu2}, while quasiparticles in a flowing condensate have
not received much attention \cite{Woo}.

In this Letter, we study the quasiparticle dynamics in a BEC superfluid with
a vortex, a typical flowing condensate. We adopt a semiclassical approach by
following the center of mass motion of a Bogoliubov wavepacket, finding
important Berry phase effects \cite{Berry} due to the background flow.
Unlike the case with a static BEC, the quasiparticle position and momentum
are no longer canonical variables , and there are also important
modifications to the quasiparticle statistics and thermodynamic properties
of the condensates. We use these results to calculate the momentum
circulation around a vortex in an infinite superfluid and find the total
transverse force acting on the vortex is proportional to the superfluid
density \cite{Thouless}, which rules out the existence of the Iordanskii
force that is supposed to originate in the asymmetric scattering of
quasiparticles by the vortex \cite{Donnelly,Sonin,Stone}. Finally, we
propose an experimental setup to directly observe Berry phase effects
through measuring local thermal atoms momentum distribution around a vortex.

\emph{Wavepacket dynamics ---} For simplicity, we present the theory here
for a single vortex, although it is applicable to a general flowing
condensate such as vortex lattices. The dynamics of the condensate are
described by the Gross-Pitaveskii equation \cite{Ghosh} 
\begin{equation}
i\frac{\partial \psi _{0}}{\partial t}=\left( \frac{\mathbf{p}^{2}}{2}%
+g\left\vert \psi _{0}\right\vert ^{2}+V\left( \mathbf{r}\right) -\mathbf{%
\Omega \cdot }\left( \mathbf{r}\times \mathbf{p}\right) \right) \psi _{0},
\label{G-P}
\end{equation}%
where for simpler notation we have used units such that $\hbar $ and the
atomic mass $m$ are both unity, $g$ is the inter-atom coupling constant, and 
$V\left( \mathbf{r}\right) $ is the trapping potential. We have also
included an angular momentum term to take into account the possibility of
using a rotating reference frame, with $\mathbf{\Omega }$ being its angular
frequency. The condensate wavefunction may be written as $\psi _{0}=\sqrt{%
n\left( \mathbf{r}\right) }e^{i\alpha \left( \mathbf{r}\right) }e^{-i\mu t}$%
, where $\mu $ is the chemical potential, $n\left( \mathbf{r}\right) $ and $%
\alpha \left( \mathbf{r}\right) $ are the vortex density and phase
respectively. The gradient of phase yields local velocity of the vortex $%
\mathbf{v}_{s}=\nabla \alpha \left( \mathbf{r}\right) $.

Small deviation $\delta \psi e^{-i\mu t}$ (elementary excitations) from the
condensate wavefunction $\psi _{0}$ is described by the Bogoliubov equation 
\cite{Ghosh} 
\begin{equation}
i\sigma _{z}{\frac{\partial }{\partial t}}\left\vert \Phi \right\rangle =%
\mathcal{Q}\left\vert \Phi \right\rangle ,\text{ }\mathcal{Q}=\left( \!%
\begin{tabular}{cc}
$H_{+}$ \space & $H_{2}e^{2i\alpha \left( \mathbf{r}\right) }$ \\ 
$H_{2}e^{-2i\alpha \left( \mathbf{r}\right) }$ \space & $H_{-}$%
\end{tabular}%
\ \ \ \!\right) \ .  \label{Bogo}
\end{equation}%
The wavefunction $\Phi $ has two components, $u$ and $v$, which are related
to $\delta \psi $ through $\delta \psi =ue^{-i\omega t}+v^{\ast }e^{i\omega
t}$, where $\omega $ is the quasiparticle energy. The entries of the matrix
operator are given by $H_{\pm }=\mathbf{p}^{2}/2+2gn\left( \mathbf{r}\right)
+V\left( \mathbf{r}\right) -\mu \mp \mathbf{\Omega \cdot }\left( \mathbf{r}%
\times \mathbf{p}\right) $, and $H_{2}\left( \mathbf{r}\right) =gn\left( 
\mathbf{r}\right) $. Because of the Pauli matrix $\sigma _{z}$ on the left
hand side, the Bogoliubov equation is nonhermitian.

We now consider a quasiparticle wavepacket centered at $\mathbf{r}_{c}$ with
its spread small compared to the length scale of the slowly varying
potentials (including trap potential $V\left( \mathbf{r}\right) $,
condensate wavefunction $\psi _{0}$ and terms related to $\mathbf{\Omega }$%
). The dynamics of the wavepacket is approximately governed by the local
Bogoliubov operator $\mathcal{Q}_{c}\equiv \mathcal{Q}\left( \mathbf{p},%
\mathbf{r}=\mathbf{r}_{c}\right) $ plus its linear gradient correction. The
local Bogoliubov operator has plane wave eigenstates $\mathcal{Q}_{c}e^{i%
\mathbf{q}\cdot \mathbf{r}}\left\vert \phi \left( \mathbf{q,r}_{c}\right)
\right\rangle =\omega _{c}\sigma _{z}e^{i\mathbf{q}\cdot \mathbf{r}%
}\left\vert \phi \left( \mathbf{q,r}_{c}\right) \right\rangle $, where $%
\mathbf{q}$ is the wavevector. The amplitude satisfies a simple 2x2 matrix
equation, which can be solved easily to yield 
\begin{equation}
\omega _{c}\left( \mathbf{q},\mathbf{r}_{c}\right) =\left( H_{1}^{2}\left( 
\mathbf{q}\right) -H_{2}^{2}\right) ^{1/2}-\mathbf{q}\cdot \left( \mathbf{%
\Omega \times r}_{c}\right)  \label{Localene}
\end{equation}%
for the local quasiparticle energy, where $H_{1}\left( \mathbf{q}\right) =%
\mathbf{q}^{2}/2+2gn\left( \mathbf{r}_{c}\right) +V\left( \mathbf{r}%
_{c}\right) -\mu $, and $\left\vert \phi \left( \mathbf{q,r}_{c}\right)
\right\rangle =\frac{1}{2}\left( \!%
\begin{array}{c}
\zeta +\zeta ^{-1} \\ 
\left( \zeta -\zeta ^{-1}\right) e^{-2i\alpha \left( \mathbf{r}_{c}\right) }%
\end{array}%
\!\right) $ for the two-component amplitude of the local eigenstate, where $%
\zeta =\left( \frac{H_{1}-H_{2}}{H_{1}+H_{2}}\right) ^{1/4}$. We note that
the wavevector $\mathbf{q}$ and the wavepacket center $\mathbf{r}_{c}$ enter
the local quasiparticle energy and wavefunction parametrically. The
wavefunction is normalized in the sense that $\left\langle \phi \left( 
\mathbf{q,r}_{c}\right) \right\vert \sigma _{z}\left\vert \phi \left( 
\mathbf{q,r}_{c}\right) \right\rangle =1$. We have also chosen the phase of
the wavefunction such that it is smooth and single valued in the parameters $%
(\mathbf{q},\mathbf{r}_{c})$. We shall see that the parametric dependence of
the eigenstates on the center position of the wavepacket will manifest as
Berry-phase terms in the equations of motion.

We now turn our attention to the wavepacket itself, which is to be
constructed out of these eigenstates as 
\begin{equation}
\left\vert \Phi \right\rangle =\int d^{3}qa\left( \mathbf{q},t\right) e^{i%
\mathbf{q}\cdot \mathbf{r}}\left\vert \phi \left( \mathbf{q,r}_{c}\right)
\right\rangle ,  \label{wavepack}
\end{equation}%
where the superposition amplitude $a\left( \mathbf{q},t\right) $ may be
taken as a Gaussian in $\mathbf{q}$. The normalization is taken to be $%
\left\langle \Phi \right\vert \sigma _{z}\left\vert \Phi \right\rangle =\int
d^{3}q\left\vert a\left( \mathbf{q},t\right) \right\vert ^{2}=1$. We assume
that the Gaussian is centered at $\mathbf{q}_{c}=\int d^{3}q\left\vert
a\left( \mathbf{q},t\right) \right\vert ^{2}\mathbf{q}$ and has a width
narrow compared to momentum scales of the energy dispersion and of the
eigenstates. Microscopic calculation shows that $\mathbf{q}_{c}=\left\langle
\Phi \right\vert \sigma _{z}\mathbf{p}\left\vert \Phi \right\rangle =\int
\delta \psi ^{\ast }\mathbf{p}\delta \psi d\mathbf{r}$, implying that it
represents the mechanical momentum of the quasiparticle. To be self
consistent, the wavepacket must yield the preassigned center position $%
\mathbf{r}_{c}=\left\langle \Phi |\sigma _{z}\mathbf{r}|\Phi \right\rangle $.

The dynamics of the quasiparticle can be derived from a time-dependent
variational principle for the Bogoliubov equation, where the action (defined
as time integral of the Lagragian) is extremized with respect to the
quasiparticle wavefunction. Here the Lagrangian is given by 
\begin{equation}
L=-\left\langle \Phi \right\vert \mathcal{Q}\left\vert \Phi \right\rangle
+\left\langle \Phi \right\vert i\sigma _{z}\frac{d}{dt}\left\vert \Phi
\right\rangle ,  \label{Lagr}
\end{equation}%
and the wavepacket (\ref{wavepack}) is chosen as the variational
wavefunction with time-dependent parameters $\mathbf{r}_{c}$ and $\mathbf{q}%
_{c}$. We use ${d}/{dt}$ to mean the derivative with respect to the time
dependence of the wavefunction explicitly or implicitly through $\mathbf{r}%
_{c}$ and $\mathbf{q}_{c}$. Under the previously discussed conditions that
the wavepacket is narrow both in position and momentum spaces, the
Lagrangian (\ref{Lagr}) can be evaluated as a function of variational
parameters $\mathbf{r}_{c}$ and $\mathbf{q}_{c}$, and their time
derivatives, independent of the width and shape of the wavepacket in
position or momentum.

\emph{Quasiparticle energy\ and Berry phase ---} The first term in
Lagrangian (\ref{Lagr}) corresponds to the total energy of the quasiparticle
wavepacket, which is found to be 
\begin{equation}
\omega =\left[ H_{1}^{2}\left( \mathbf{q}_{c}\right) -H_{2}^{2}\right]
^{1/2}-\mathbf{q}_{c}\cdot \left( \mathbf{\Omega \times r}_{c}\right)
+\left( 1-\rho ^{2}\right) \mathbf{q}_{c}\cdot \mathbf{v}_{s}.  \label{Ene1}
\end{equation}%
where $\mathbf{v}_{s}=\nabla \alpha \left( \mathbf{r}_{c}\right) $ is the
local velocity of the vortex. The first two terms are the local
quasiparticle energy Eq.(\ref{Localene}) with momentum $\mathbf{q}_{c}$,
stemming from the expectation value $\left\langle \Phi \right\vert \mathcal{Q%
}_{c}\left\vert \Phi \right\rangle $ of the local Bogoliubov operator. The
last term is the correction of the wavepacket energy originating from the
linear gradient expansion of the Bogoliubov operator $\Delta \mathcal{Q}=%
\frac{1}{2}\left( \left( \mathbf{r}-\mathbf{r}_{c}\right) \cdot \frac{%
\partial \mathcal{Q}_{c}}{\partial \mathbf{r}_{c}}+c.c\right) $, and $\rho
=\left\langle \Phi |\Phi \right\rangle $ is the total atomic mass contained
in the quasiparticle wavepacket.

To understand the quasiparticle energy expression, it will be instructive to
consider the simpler situation of a uniformly flowing superfluid of velocity 
$\mathbf{v}_{s}$. If $\mathbf{p}_{0}$ is the quasiparticle momentum in the
reference frame where the superfluid is static, then we have the
relationship $\mathbf{q}_{c}=\mathbf{p}_{0}+\rho \mathbf{v}_{s}$. In terms
of $\mathbf{p}_{0}$, the quasiparticle energy (\ref{Ene1}) can be expressed
(in the limit of small $\mathbf{v}_{s}$) as $\omega =\varepsilon \left( 
\mathbf{p}_{0}\right) +\mathbf{p}_{0}\cdot \mathbf{v}_{s}$, where $%
\varepsilon \left( \mathbf{p}_{0}\right) =\left[ H_{1}^{2}\left( \mathbf{p}%
_{0}\right) -H_{2}^{2}\right] ^{1/2}=\left[ p_{0}^{2}\left(
p_{0}^{2}/4+gn\right) \right] ^{1/2}$ is the energy dispersion in a static
superfluid. This is just the standard Landau formula obtained using Galilean
transformation \cite{Landau}.

The evaluation of the second term of the Lagrangian (\ref{Lagr}) is similar
to that in Ref. \cite{Niu1}, which yields 
\begin{equation}
L=-\omega +\left( \mathbf{q}_{c}+\mathbf{A}\right) \cdot \mathbf{\dot{r}}%
_{c}.  \label{Lagr2}
\end{equation}%
Here the term $\mathbf{q}_{c}\cdot \mathbf{\dot{r}}_{c}$ comes from the
time-dependence of the superposition amplitude $a\left( \mathbf{q},t\right) $
in the wavepacket (\ref{wavepack}), while the vector potential $\mathbf{A}%
=i\left\langle \phi \right\vert \sigma _{z}\left\vert \partial \phi
/\partial \mathbf{r}_{c}\right\rangle =-(\rho -1)\nabla \alpha \left( 
\mathbf{r}_{c}\right) $ is obtained from the position dependence of the
two-component local eigenvector, which is similar to the case of a spin in a
position-dependent Zeeman field \cite{Berry}. The vector potential, also
called Berry connection, arises for a vortex because of the non-zero local
velocity. Its line integral over a path $\mathcal{C}$ gives a Berry phase $%
\Gamma \left( \mathcal{C}\right) =\oint_{\mathcal{C}}d\mathbf{r}_{c}\cdot 
\mathbf{A}=-\oint_{\mathcal{C}}\left( \rho -1\right) d\alpha \left( \mathbf{r%
}_{c}\right) $ of the eigenvector \cite{Stoof}. For instance, the
accumulated Berry phase for a quasiparticle moving around a vortex in an
infinite uniform superfluid is $-2\pi \left( \rho -1\right) $, where $\rho $
represents the total atom number (notice that the atom mass $m$ has been
taken as unity) in the quasiparticle wavepacket. Apart from a non-essential $%
2\pi $, it means that the phase acquired by moving each atom around a vortex
is $2\pi $, which agrees with Feynman's argument \cite{Feynman}.

We note that the vector potential vanishes at very high momenta, where $\rho
\rightarrow 1$. Likewise, the energy correction due to superfluid flow (the
last term of Eq.(\ref{Ene1})) vanishes in this limit. This is quite
reasonable, because the quasiparticle becomes a free particle and is
decoupled from the condensate at high momenta. The condensate has influence
on the quasiparticle only at low and intermediate momenta.

\emph{Semiclassical dynamics ---} Following the standard procedure of
analytical mechanics, we may introduce the canonical momentum $\mathbf{k}%
_{c}=\partial L/\partial \mathbf{\dot{r}}_{c}=\mathbf{q}_{c}+\mathbf{A}$
conjugate to the coordinate vector $\mathbf{r}_{c}$. We see that the vector
potential makes the canonical and mechanical momenta different. The energy $%
\omega =\mathbf{k}_{c}\cdot $ $\mathbf{\dot{r}}_{c}-L$ becomes the
Hamiltonian and the equations of motion are 
\begin{equation}
\mathbf{\dot{r}}_{c}=\frac{\partial \omega }{\partial \mathbf{k}_{c}},%
\mathbf{\dot{k}}_{c}=-\frac{\partial \omega }{\partial \mathbf{r}_{c}},
\label{EM1}
\end{equation}%
where the quasiparticle energy $\omega $ is now recasted as a function of $%
\mathbf{k}_{c}$, $\omega =\left[ H_{1}^{2}\left( \mathbf{k}_{c}\right)
-H_{2}^{2}\right] ^{\frac{1}{2}}-\left( \mathbf{k}_{c}-\mathbf{A}\right)
\cdot \left( \mathbf{\Omega }\times \mathbf{r}_{c}\right) +\mathbf{k}%
_{c}\cdot \mathbf{A}$. We see that the vector potential $\mathbf{A}$
modifies the energy expression in this canonical formulation.

In terms of the physical quantity of mechanical momentum $\mathbf{q}_{c}$,
the quasiparticle equations of motion are dramatically altered: 
\begin{eqnarray}
\mathbf{\dot{r}}_{c} &=&\frac{\partial \omega }{\partial \mathbf{q}_{c}}+%
\frac{\partial \rho }{\partial \mathbf{q}_{c}}\left( \mathbf{\dot{r}}%
_{c}\cdot \mathbf{v}_{s}\right) ,  \label{EM2} \\
\mathbf{\dot{q}}_{c} &=&-\frac{\partial \omega }{\partial \mathbf{r}_{c}}-%
\mathbf{\dot{r}}_{c}\times \left( \frac{\partial \rho }{\partial \mathbf{r}%
_{c}}\times \mathbf{v}_{s}\right) +\left( \mathbf{\dot{q}}_{c}\cdot \frac{%
\partial \rho }{\partial \mathbf{q}_{c}}\right) \mathbf{v}_{s}\text{.} 
\nonumber
\end{eqnarray}%
They follow directly from (\ref{EM1}) by the transformation $\mathbf{k}_{c}=%
\mathbf{q}_{c}+\left( 1-\rho \right) \mathbf{v}_{s}$. Here $\omega $ is the
original energy expression (\ref{Ene1}). We see that additional terms
depending on superfluid velocity and total atom mass in the quasiparticle
wavepacket appear in both equations, therefore the equations of motion are
no longer of canonical form. This a result very different from the case of
static superfluid studied in \cite{Csordas}.

In statistical treatment of quasiparticles, an important semiclassical
quantity is the density of states of phase space. For the canonical
variables $\mathbf{k}_{c}$ and $\mathbf{r}_{c}$, this should be taken as the
constant $1/\left( 2\pi \right) ^{3}$. Because of the vector potential, the
position $\mathbf{r}_{c}$ and mechanical momentum $\mathbf{q}_{c}$ are no
longer canonical variables, which leads to a modification of the density of
states \cite{Xiao} 
\begin{equation}
D\left( \mathbf{r}_{c},\mathbf{q}_{c}\right) =\frac{1}{\left( 2\pi \right)
^{3}}\det \left( \mathbf{I}-\frac{\partial \rho }{\partial \mathbf{q}_{c}}%
\mathbf{v}_{s}\right) ,  \label{DS}
\end{equation}%
where $\mathbf{I}$ is the 3$\times $3 unit matrix.

The thermodynamics properties of condensates are determined by the physical
quantities of quasiparticles such as energy and density of states, therefore
they are strongly affected by Berry phase. For instance, the thermal
depletion of the condensate density is given by 
\begin{equation}
\delta n\left( \mathbf{r}_{c},T\right) =\int d^{3}\mathbf{q}_{c}D\left( 
\mathbf{r}_{c},\mathbf{q}_{c}\right) \rho /\left( e^{\omega
/k_{B}T}-1\right)   \label{deltaN2}
\end{equation}%
according to our theory, which is different from that for a static
condensate because of the modifications of energy $\omega $ and density of
state $D\left( \mathbf{r}_{c},\mathbf{q}_{c}\right) $. Notice that the
integration of this distribution over position space yields the total number
of thermal atoms.

\emph{Transverse force on a vortex ---} The general form of the transverse
force per unit length acting on a moving vortex with velocity $\mathbf{v}%
_{L} $ in an infinite uniform superfluid can be written as $\mathbf{F=\kappa
\left( \mathcal{C}\right) \hat{z}\times v}_{L}$, where $\mathbf{\kappa
\left( \mathcal{C}\right) }$ is the momentum circulation along a path far
away from the core of the vortex, and we assume the normal fluid velocity is
zero \cite{Stone}. In the past several decades, there has been some
controversy about the expression of $\mathbf{\kappa \left( \mathcal{C}%
\right) }$, which was argued to be either $2\pi \hbar n_{tot}/m$ or $2\pi
\hbar n_{s}/m$ through different approaches \cite{Thouless,Donnelly,Sonin}.
Here $2\pi \hbar /m$ is the quantum of circulation, $n_{tot}$ is the total
mass density, and $n_{s}$ is the superfluid density .

According to our theory, the total momentum circulation $\mathbf{\kappa
\left( \mathcal{C}\right) }$ contains two contributions: $2\pi \hbar \left(
n_{tot}-\delta n\left( \mathbf{r}_{c},T\right) \right) /m$ from the
condensate and $\mathbf{\kappa }_{1}\left( \mathbf{\mathcal{C}}\right)
=\oint_{\mathcal{C}}\mathbf{W}\cdot d\mathbf{r}_{c}$ from quasiparticles. $%
\mathbf{W}$ is the momentum density of quasiparticles under equilibrium
Bose-Einstein distribution and is obtained by summing over the mechanical
momenta from all states, 
\begin{eqnarray}
\mathbf{W} &=&\int d^{3}\mathbf{q}_{c}D\left( \mathbf{r}_{c},\mathbf{q}%
_{c}\right) \frac{\mathbf{q}_{c}}{e^{\omega \left( \mathbf{r}_{c},\mathbf{q}%
_{c}\right) /k_{B}T}-1}  \label{momentden} \\
&=&\int \frac{d^{3}\mathbf{k}_{c}}{\left( 2\pi \right) ^{3}}\frac{\mathbf{k}%
_{c}+\left( \rho -1\right) \nabla \theta _{c}}{e^{\omega \left( \mathbf{r}%
_{c},\mathbf{k}_{c}\right) /k_{B}T}-1},  \nonumber
\end{eqnarray}%
where $T$ is the temperature, $k_{B}$ is the Boltzman constant, $D\left( 
\mathbf{r}_{c},\mathbf{q}_{c}\right) =\frac{1}{\left( 2\pi \right) ^{3}}%
\left( 1-\frac{1-\rho ^{2}}{\omega _{c}r_{c}^{2}}\left( \mathbf{r}_{c}\times 
\mathbf{q}_{c}\right) _{z}\right) $, and $1/\left( 2\pi \right) ^{3}$ are
the density of states for the non-canonical momentum $\mathbf{q}_{c}$ and
canonical momentum $\mathbf{k}_{c}$ respectively.

Since we choose the circular contour $\mathcal{C}$ to be far away from the
core of the vortex, we can expand the Bose-Einstein distribution $f\left( 
\mathbf{k}_{c}\right) \approx f_{0}\left( \mathbf{k}_{c}\right) -\Delta f$
with $f_{0}\left( \mathbf{k}_{c}\right) =\frac{1}{e^{\omega _{c}/k_{B}T}-1}$%
, and $\Delta f=\frac{e^{\omega _{c}/k_{B}T}}{\left( e^{\omega
_{c}/k_{B}T}-1\right) ^{2}}\frac{\left( 1-\rho \right) \mathbf{k}_{c}\mathbf{%
\cdot }\nabla \theta _{c}}{k_{B}T}$. Considering a closed orbit with fixed
radius $R$ and taking the limit of $R\rightarrow \infty $, we find the
thermal depletion of the condensate $\delta n\left( \mathbf{r}_{c},T\right) =%
\frac{1}{2\pi ^{2}}\int_{0}^{\infty }f_{0}\left( \mathbf{k}_{c}\right) \rho
k_{c}^{2}dk_{c}$, and quasiparticle momentum circulation%
\begin{eqnarray}
\kappa _{1}\left( \mathcal{C}\right) &=&\frac{1}{\pi }\int_{0}^{\infty
}f_{0}\left( \mathbf{k}_{c}\right) \rho k_{c}^{2}dk_{c}  \label{cir} \\
&&-\frac{1}{3\pi k_{B}T}\int_{0}^{\infty }\frac{e^{\omega
_{c}/k_{B}T}k_{c}^{4}dk_{c}}{\left( e^{\omega _{c}/k_{B}T}-1\right) ^{2}}. 
\nonumber
\end{eqnarray}%
The first term of $\kappa _{1}\left( \mathcal{C}\right) $ stems from the
difference between the canonical momentum $\mathbf{k}_{c}$ and mechanical
momentum $\mathbf{q}_{c}$, and cancels with $-2\pi \hbar \delta n\left( 
\mathbf{r}_{c},T\right) /m$. The second term of $\kappa _{1}\left( \mathcal{C%
}\right) $ comes from the deviation $\Delta f$ in the distribution function,
and is found to be $2\pi \hbar \rho _{n}/m$, where $\rho _{n}=\frac{2\pi ^{2}%
}{45}\frac{\left( k_{B}T\right) ^{4}}{\hbar ^{3}s^{5}}$ is the normal fluid
density. Summing up the contributions from both condensate and
quasiparticles, we find the total momentum circulation $\kappa \left( 
\mathcal{C}\right) =2\pi \hbar \left( n_{tot}-\rho _{n}\right) /m=2\pi \hbar
n_{s}/m$, and the transverse force per unit length $\mathbf{F}=2\pi \hbar
n_{s}\mathbf{\hat{z}}\times \mathbf{v}_{L}/m$\textbf{. }This form of
transverse force agrees with that obtained by Thouless \textit{et. al. }%
using general properties of superfluid order \cite{Thouless}, and rules out
the existence of the Iordanskii force that is proportional to the normal
fluid density $\rho _{n}$, and is supposed to originate in the asymmetric
scattering of quasiparticles by the vortex \cite{Donnelly,Sonin,Stone}.

\emph{Experimental observation --- } The above discussions show that the
modifications of mechanical momentum, energy dispersion and density of
states of quasiparticles due to Berry phase affect the thermodynamics of
atoms. Our prediction of a transverse force proportional $T^{4}$ can in
principle be measured in experiments. In the following, we propose an
experiment for a direct observation of the local thermal atom momentum
distribution around a vortex, in which the effect of Berry phase can be seen
clearly. 
\begin{figure}[t]
\begin{center}
\vspace*{-0.0cm}
\par
\resizebox *{8cm}{6cm}{\includegraphics*{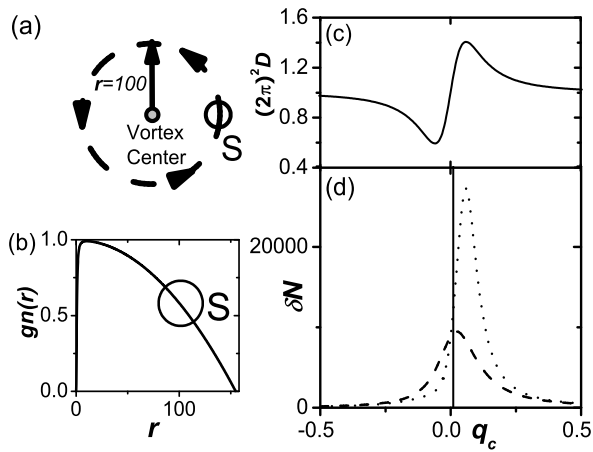}}
\end{center}
\par
\vspace*{0.0cm}
\caption{Thermal-atom momentum distribution around a trapped vortex. Length
scales are in units of $l=0.73$ $\protect\mu $m for a Rb$^{87}$ BEC in a
trap specified in the text. (a) A schematic plot of experimental geometry.
Measurement of thermal-atom momentum distribution is to be made in a region $%
S$\ at $100$ unit lengths away from the vortex center with a radius of 20
unit lengths. This region contains about $9740$ condensate and $1650$
thermal atoms. (b) Condensate atom density (multiplied by the interaction
strength $g$) in the presence of a trapped vortex. (c) Density of states for
quasiparticles in the measurement region $S$ as a function of quasiparticle
momentum $q_{c}$ along the circulation flow. (d) Momentum distribution of
thermal atoms in the measurement region $S$. Dotted and dashed lines
correspond to our Eq. (\protect\ref{deltaN3}) and the naive expression $%
\protect\int_{S}d\mathbf{r}_{c}\protect\rho /\left( \exp \left( \frac{%
\protect\omega _{c}+\mathbf{q}_{c}\cdot \protect\nabla \protect\theta }{%
k_{B}T}\right) -1\right) $, respectively. For reference, the momentum of
condensate atoms is indicated by the vertical solid line.}
\label{rr}
\end{figure}

Fig.1(a) is a schematic plot of experimental geometry. The length unit is $l=%
\sqrt{\frac{\hbar }{m\left( \mu +\hbar \Omega \right) }}$, chosen for
convenience. It corresponds to $l\approx 0.73$ $\mu m$ for a BEC of $%
N=5\times 10^{5}$ Rb$^{87}$ atoms in a quasi-two dimensional magnetic trap 
\cite{Ketterle} with axial and radial trapping frequencies $\omega _{z}=2\pi
\times 800$ Hz, $\omega _{r}=2\pi \times 2$ Hz, in which a single vortex is
created at a rotation frequency of $\Omega =0.4\omega _{r}$ \cite{Matthews}.
The condensate density profile is shown in Fig.1(b), where one can see that
the observation region $S$ is chosen relatively far away from the vortex
core, which was to make sure that the semiclassical approximation remains
valid. We will consider a temperature of $T=2\mu /k_{B}\approx 21.1$ nK
which is about a third of the BEC transition temperature $T_{c}\approx 52.8$
nK for this system.

The atoms are supposed to be in a hyperfine state ($\left\vert
1\right\rangle $). By focusing two copropagating Raman beams at region $S$
one may drive both the thermal and condensate atoms there to another
hyperfine state ($\left\vert 2\right\rangle $). The remaining atoms in state 
$\left\vert 1\right\rangle $ are to be rapidly expelled from the trap by
applying radio-frequency radiation that flips them to an anti-trapped
hyperfine state. Finally, the time of flight measurement \cite{Raizen} of
atoms in state $\left\vert 2\right\rangle $ can then determine their
momentum distribution, which, according to our theory, is given by 
\begin{equation}
\delta n\left( \mathbf{p}_{c},T\right) =\int_{S}d^{2}\mathbf{r}_{c}D\left( 
\mathbf{r}_{c},\mathbf{q}_{c}\right) \rho /\left( e^{\omega
/k_{B}T}-1\right) .  \label{deltaN3}
\end{equation}%
In Fig.1(c), we see that there are more quasiparticles moving along the
circulating superfluid flow than in the opposite direction and the deviation
from the naive value of the density of states $1/\left( 2\pi \right) ^{2}$
is substantial. In Fig.1(d), we find that the momentum distribution is quite
different from the commonly believed one that does not take into count the
Berry phase effects.

We gratefully thank Junren Shi, Alexander L. Fetter, and Erich Mueller for
stimulating discussions. This work is supported by the NSF and the R.A.
Welch foundation.

\end{document}